 \documentclass[arguments]{aastex63}
\accepted{\today}
\submitjournal{ApJ}

\shorttitle{A disk-outflow model for CL AGNs}
\shortauthors{Feng et al.}
\graphicspath{{./}{figures/}}

\begin{document}

\title{A magnetic accretion disk-outflow model for changing look active galactic nuclei}

\correspondingauthor{Xinwu Cao}

\author{Junjie Feng}
\affiliation{Key Laboratory for Research in Galaxies and Cosmology, Shanghai Astronomical Observatory, Chinese Academy of Sciences, 80 Nandan Road, Shanghai, 200030, China}
\affiliation{University of Chinese Academy of Sciences, 19A Yuquan Road, 100049, Beijing, China}

\author{Xinwu Cao}
\affiliation{Zhejiang Institute of Modern Physics, Department of Physics, Zhejiang University, 38 Zheda Road, Hangzhou 310027, China, Email: xwcao@zju.edu.cn}
\affiliation{Shanghai Astronomical Observatory, Chinese Academy of Sciences, 80 Nandan Road, Shanghai, 200030, China}
\affiliation{ Key Laboratory of Radio Astronomy, Chinese Academy of Sciences,
	210008 Nanjing, China}
\author{Jia-wen Li}

\affiliation{Zhejiang Institute of Modern Physics, Department of Physics, Zhejiang University,
38 Zheda Road, Hangzhou 310027, China }

\author{Wei-Min Gu}

\affiliation{Department of Astronomy, Xiamen University, Xiamen, Fujian 361005, China}




\begin{abstract}
The time-scales of the variabilities in changing look (CL) active galactic nuclei (AGNs) are usually at the order of years to tens of years (some of them are even shorter than one year), which are much shorter than the viscous timescale of a standard thin accretion disk. It implies that the variabilities of CL AGNs cannot be reproduced by varying the mass accretion rate of the thin disk. In this work, we employ a magnetic accretion disk-outflow model to calculate the inflow time of the disk predominantly driven by magnetic outflows. In this model, most angular momentum of the gas in the disk is carried away by the outflows, and therefore its radial velocity can be substantially higher than that of a conventional viscous disk. Our calculations show that the inflow time of such a disk with outflows can be around several years to tens years. The calculated spectra of the disk with outflows can fit the observed spectra of a CL AGN Mrk 1018 quite well both in the low and high states. The derived inflow time of such a disk with outflows is around 5 years in the high state, while it becomes $\sim 20$ years in the low state, which is roughly consistent with the observations of the variabilities in Mrk 1018.
\end{abstract}



\section{Introduction} \label{sec:intro}
The accretion onto supermassive black hole (SMBH) in the center of galaxies is the dominant energy source of the observed active galactic nuclei (AGNs). The AGNs can be roughly classified by the width of the emission lines, range from Seyfert 1 to Seyfert 2 \citep{1943ApJ....97...28S}. As the classical unified model of AGNs proposed, the type of an AGN is quite definite, which is usually thought to be dependent on the line of observers' sight   \citep{1989agna.book.....O}. However, many changing-look AGNs (CL AGNs) have been discovered in the past decades, which is a term describing the AGNs changing from Seyfert 1 to Seyfert 2 or vice versa, by the sudden brighten or dimming during a short period. With an increasing number of multi-band surveys, the number of CL AGNs is growing rapidly  \citep[e.g.][]{1976ApJ...210L.117T,2016A&A...593L...8M,2018ApJ...862..109Y,2019ApJ...883L..44G,2019ApJ...883...94T,2020AJ....159..245W}, however, the physics responsible for such violent variabilities is still a mystery.

There are several popular scenarios employed to explain the observational features of CL AGNs. One scenario is that the broad emission lines are obscured by the torus or moving clouds over observer's line of sight \citep{1989ApJ...346L..21G}, while only several CL AGNs can be explained by this scenario \citep{2019ApJ...887...15W,2020MNRAS.493..930J,2020MNRAS.491.4615K}. The features observed in most CL AGNs, e.g., the complex multi-band spectral variabilities, and strong changes seen in the infrared or low level of polarization, strongly argue against the scattering (or obscuration) scenario \citep[e.g.][and the references therein]{2019MNRAS.485.2573K,2018ApJ...866..123M,2017ApJ...846L...7S,2018ApJ...864...27S,2019A&A...625A..54H}. Another attractive scenario is that the CL AGNs are indeed tidal disruption events (TDEs) \citep{2015MNRAS.452...69M,2016PASJ...68...58K,2019ApJ...885..110Y,2020ApJ...898L...1R,2020arXiv200307365P,2021MNRAS.500L..57Z}, while it is not a general mechanism to explain all CL AGNs, such as repeating CL AGNs.
The changing look phenomenon triggered by the change of the mass accretion rate of the accretion disk is a rather straightforward model \citep{2016A&A...593L...9H,2019ApJ...883...76R,2019arXiv191203972L,2020ApJ...890L..29A,2020A&A...641A.167S},  however, it has a fatal problem that propagation time-scale of the gas in a thin accretion disk is much longer than observed time-scale in CL AGNs, unless the viscous thin disk model is somewhat revised \citep{2018NatAs...2..102L}. \citet{2019MNRAS.483L..17D} proposed that the magnetically elevated disk model could help to explain the changing look time-scale. Based on this scenario, \citet{2021MNRAS.502L..50S} suggested that a magnetic flux inversion in a magnetically arrested disk is able to explain the CL event in 1ES 1927+654.  \citet{2020A&A...641A.167S} suggested that a narrow unstable zone between the outer thin disk and the inner ADAF could cause the periodic outbursts in repeating CL AGNs. Recently, the effects of large-scale magnetic fields on this scenario for repeating CL AGNs have been studied in detail by \citet{2021ApJ...910...97P}.

It is well-known that the outflows have been observed in many AGNs   \citep{2009MNRAS.397.1836G,2011MNRAS.417..464M,2015ApJ...806...22D,2020MNRAS.495..305X,2020MNRAS.494.5396B}, and the formation and acceleration of outflows have also been extensively studied by the numerical simulations \citep{2014ApJ...796..106J,2015ApJ...804..101Y,2017MNRAS.465.2873N,2018ApJ...867..100Y,2020MNRAS.497.5229C} and theoretical works \citep{1982MNRAS.199..883B,1994MNRAS.268.1010L,2010A&A...516A..89R,2014ApJ...783...51C,2015MNRAS.448.3514C,2019ApJ...885...93F}.
Magnetically driven outflows is one of most promising mechanism \citep*[][]{1982MNRAS.199..883B}, in which a large-scale magnetic fields is a necessary ingredient. It has been suggested that the external weak magnetic fields can be transported inwards by the accretion flow, while it encounters difficulty in geometrically thin disk due to its small radial velocity of the gas in the disk  \citep*[][]{1994MNRAS.267..235L,2005ApJ...629..960S}. \citet{2013ApJ...765..149C} suggested that the external field can be efficiently advected inwards by a thin disk if most angular momentum of the disk is carried away by the magnetic outflows.
\citet{2019ApJ...872..149L} derived a global solution to a thin disk with magnetic driven outflows, in which the radial velocity is increased to tens or even hundreds times higher than that of a conventional viscous thin disk.
It implies that the inflow time of such a disk with outflows can be much shorter than that of a standard thin disk. In this work, we apply the accretion disk-outflow model developed by \citet{2019ApJ...872..149L} to explain the observational features of CL AGNs. We describe the magnetic accretion disk-outflow model in Section \ref{sec:methods}. The numerical method and the results are given in Section \ref{sec:results}-\ref{sec:results_b}. Section \ref{sec:discussion} contains our discussion and conclusions.

\section{Model} \label{sec:methods}

For a standard thin disk, the viscous time-scale is
\begin{equation}
    \tau_{\rm vis}\sim -\frac{R}{v_{\rm {R, \rm vis}}}. \label{time}
\end{equation}
Assuming the optical/UV photons are emitted at $R \sim 50R_{\rm s}$, the viscous time-scale is about $10^5$ years for $\alpha = 0.1$, which is several orders of magnitude higher than the typical time-scales of variabilities observed in CL AGNs. As discussed in \citet{2019ApJ...886...92L}, in a thin disk with magnetic outflows, the magnetic torque exerted by the outflows is the dominant term to drive the mass accretion, and the angular momentum of the disk is transferred more effective with strong magnetic outflows. Thus the radial inflow time-scale will be shortened for a thin disk with strong magnetic outflows. A thin accretion disk model with magnetic outflows has been developed in the previous work  \citep[see][]{2013ApJ...765..149C,2019ApJ...872..149L,2019ApJ...886...92L}. We adopt the accretion disk-outflow model to explain the variability time-scales and the spectra of the CL AGNs. Such an accretion disk with magnetically driven outflows is described by a set of  equations of the disk and outflows. In this work, we summarize the model briefly in the following sub-sections \citep[see][for the details of the model]{2013ApJ...765..149C,2019ApJ...872..149L,2019ApJ...886...92L}. We note that a series of works based on similar accretion disk-outflow coupling scenario have successfully explain the state transitions in the galactic black hole X-ray binaries \citep{2006A&A...447..813F,2018A&A...615A..57M,2018A&A...617A..46M,2019A&A...626A.115M,2020A&A...640A..18M}, which are somewhat similar to the observational features in some CL AGNs.

\subsection{Structure of the disk with magnetically driven outflows}\label{disk_structure}

In the case of a disk with outflows, the mass accretion rate
\begin{equation}
\dot{M} (R) = -2\pi R\Sigma v_{R},\label{mdot}
\end{equation}
is no longer a constant along $R$ due to mass loss in the outflows. If the mass loss rate
$\dot{m}_{\rm w}$ from the unit area (one surface) of the disk is known, the mass conversation in the disk requires
\begin{equation}
\frac{d\dot{M} }{d R} = 4\pi R\dot{m}_{\rm w}.\label{dM_acc-dR}
\end{equation}

The angular momentum equation of the disk with magnetic outflows reads
\begin{equation}\label{angular_eq}
\frac{d}{d R}\left( 2\pi R \Sigma v _{ R} R^2 \Omega \right) = \frac{d}{d R}
\left( 2 \pi R \nu \Sigma R^2 \frac{d \Omega}{d R} \right) + 2 \pi R T_{\rm m} ,
\end{equation}
where $T_{\rm m}$ is the magnetic torque exerted on the unit surface of the disk due to the outflows. Integrating Equation (\ref{angular_eq}), the radial velocity, $v _R$, of an accretion disk with magnetic outflows can be calculated as
\begin{equation}\label{v_R}
v _{ R}
=-\frac{3\alpha c_{s}H}{2R} - \frac{T_{\rm m}}{\Sigma}\left[ \frac{\partial}{\partial R}\left( R^2 \Omega\right)  \right] ^{-1}
\simeq-\frac{3\alpha c_{s}H}{2R} - \frac{2T_{\rm m}}{\Sigma R \Omega}
= v_{R,  \rm{vis}} + v_{R,\rm m},
\end{equation}
where the approximation ${\rm d}\Omega/{\rm d}R\sim -3\Omega/2R$ is adopted. We use a parameter $f_{\rm m}\equiv v_{R,\rm m}/v_{R,\rm vis}$ to describe the relative importance of the angular momentum transfer mechanisms of the disk, and the radial velocity of the gas in the disk is
\begin{equation}\label{v_vis/vm}
v _{R}=(1+f_{\rm m})v_{R,  {\rm {vis}}},
\end{equation}
where the parameter $f_{\rm m}$ is to be calculated with suitable outflow model (see Section \ref{outflow}).

In the presence of a large-scale magnetic field, a radial magnetic force is exerted on the disk against the gravity of the BH, and therefore the gas in the disk is rotating at a sub-Keplerian velocity, i.e.,
\begin{equation}\label{diff_omega}
R\left( \Omega ^2 _{\rm k} - \Omega ^2 \right) = \frac{B_{z} B_{R,\rm s}}{2\pi \Sigma},
\end{equation}
which yields
\begin{equation}\label{ome/ome_k}
f_{\Omega} \equiv \frac{\Omega}{\Omega_{\rm k}} = \left[ 1 - \frac{2R\kappa_ 0}{\beta \left( 1 + \kappa^2 _0 \right)H } \frac{c^2_{\rm{s}}}{R^2 \Omega^2 _{\rm k}}\right] ^{1/2},
\end{equation}
where $\kappa_0 =B_{z}/ B_{R, \rm s}$, and $ \beta $ is defined as
\begin{equation}\label{beta_p}
\beta  \equiv \frac{P_{\rm {gas}}+P_{\rm rad}}{P_{\rm{mag}}} ={\frac {8\pi (P_{\rm gas}+P_{\rm rad})} {B^2}}
\end{equation}
at mid-plane of the disk.

Strong large-scale magnetic field may compress the disk vertically, which has been explored in detail by \citet{2002A&A...385..289C}, however, we find that the magnetic field in our model is always not so strong to compress the disk significantly, which will be addressed in Section \ref{sec:results}. Assuming that the disk to be in vertical  equilibrium, and the scale-height of the disk is
\begin{equation}\label{H_over_R}
\frac{H}{R} = \frac{c_{\rm{s}}}{R \Omega_{\rm k}}.
\end{equation}

The viscous dissipation rate of an accretion disk with magnetically driven outflows is
\begin{equation}
Q^+_{\rm vis}=-{1\over 3}R\Sigma v_{R}(1+f_{\rm m})^{-1}\left(R \frac{d \Omega}{d R}\right)^{2}={\frac 1{6\pi}}\dot{M}(1+f_{\rm m})^{-1}\left(R \frac{d \Omega}{d R}\right)^{2}\simeq {\frac 3{8\pi}}\dot{M}(1+f_{\rm m})^{-1}f_\Omega^2\Omega_{\rm K}^2, \label{Q_plus3}
\end{equation}
and the luminosity of the disk is
\begin{equation}
L=\int 4\pi R Q^+_{\rm vis}dR={\frac 2{3}}\int\dot{M}(1+f_{\rm m})^{-1}\left(R \frac{d \Omega}{d R}\right)^{2}RdR\simeq {\frac 3{2}}\int\dot{M}(1+f_{\rm m})^{-1}
f_\Omega^2\Omega_{\rm K}^2RdR \label{lum_d}
\end{equation}
\citep*[see the detailed discussion in][]{2016ApJ...817...71C}.

For a geometrically thin accretion disk, energy advection in the disk is negligible, and we have
\begin{equation}
Q^+_{\rm vis}={\frac 3{8\pi}}\dot{M}(1+f_{\rm m})^{-1}f_\Omega^2\Omega_{\rm K}^2= Q^-_{\rm rad}=\frac{4a c T_{\rm{c}}^4}{3\tau}, \label{radiation}
\end{equation}
where the vertical optical depth $\tau=\tau_{\rm es}+\tau_{\rm ff}=\rho (\kappa_{\rm es}+\kappa_{\rm ff})H$. The equation of state is
\begin{equation}
P=P_{\rm{gas}}+P_{\rm{rad}}=\frac{k_B}{\mu_p m_p}\rho T_{\rm{c}}+\frac{1}{3}a T_{\rm{c}}^4,\label{state}
\end{equation}
and the isothermal sound speed of the disk is
\begin{equation}
c^2_{\rm{s}}=\frac{P}{\rho}=\frac{k_B}{\mu_{\rm p} m_{\rm p}} T_{\rm{c}}+\frac{1}{3\rho}a T_{\rm{c}}^4.\label{v_sound}
\end{equation}

In this work, we consider a weak external vertical homogeneous large-scale magnetic field advected inwards by an accretion disk with outflows. In the steady case, the field advection is in balance with the magnetic diffusion, i.e., the advection timescale equals to the diffusion timescale, which leads to
\begin{equation}
1+f_{\rm m}={\frac {2}{3}}{\cal{P}}_{\rm m}\kappa_0^{-1}\left({\frac {H}{R}}\right)^{-1}. \label{f_m}
\end{equation}
where the magnetic Prandtl number $ {\cal{P}}_{\rm m} = \eta/\nu $ is an input parameter of this work \citep*[see][for the detailed calculations]{2019ApJ...886...92L}.
It is around unity for turbulent disk \citep*[][]{1979cmft.book.....P,2009A&A...507...19F,2009ApJ...697.1901G}.


The surface temperature $T_{\rm s}$ of the disk is given by
\begin{equation}
T_{\rm s}=\left(\frac{Q^-_{\rm rad}}{ \sigma}\right)^{1/4}.\label{T_s}
\end{equation}
The continuum spectrum of the disk with outflows can be calculated with
\begin{equation}
    L_{\nu}=2\int \frac{h\nu^3}{c^2}\frac{2\pi RdR}{e^{h\nu/k_{\rm{B}}T_{\rm{s}}}-1}.
	\label{eq:spectra}
\end{equation}


\subsection{Magnetically driven outflows from the disk}\label{outflow}

The dynamics of the magnetically driven outflows can be calculated with the Weber-Davis model, when the configuration and strength of the large-scale magnetic field, and the boundary conditions of the outflows at the disk surface are given \citep*[][]{1967ApJ...148..217W,1994A&A...287...80C}. In the case of cold outflows, i.e., the outflows are accelerated predominantly by the magnetic field, the cold Weber-Davis model suggested by \citet{2010LNP...794..233S} is a good approximation, which is in fairly good consistent with the numerical simulations \citep*[][]{2005ApJ...630..945A}. In this work, we explore the properties of the outflows by using the cold Weber-Davis outflow model \citep[see][for details]{2010LNP...794..233S,2013ApJ...765..149C}. We summarize the model as follows.

In this model, the total magnetic torque $ T_{\rm m} $ exerting on the unit area of the disk surface can be calculated by
and the  Alfv\'en radius
\begin{equation}\label{T_m_nu}
T_{\rm m} = \frac{3}{4\pi}RB_{z} B_{\rm p}\mu \left( 1+\mu^{-2/3} \right),
\end{equation}
where $ B_{\rm p} =( B^2 _{z} + B^2 _{R,{\rm s}})^{1/2} $ is the poloidal component of the magnetic field at the disk surface. The dimensionless mass loading parameter $\mu$ is defined as
\begin{equation}
\mu = \frac{4\pi \rho _{\rm w}v_{\rm w}\Omega R}{B^2_{\rm p}}
 ={\frac{4\pi \Omega R \dot{m}_{\rm w}}{B_{z}B_{\rm p}}}, \label{mass_load_para}
\end{equation}
where $v_{\rm w}$ is the poloidal velocity of the gas in the outflow, and the mass loss flux of the outflow $ \rho _{\rm w}v_{\rm w} =  \dot{m}_{\rm w} B_{\rm p}/B_{z}$ is used, The angular velocity $ \Omega $ of the field lines co-rotating with the disk can be sub-Keplerian in the presence of the large-scale field (see the discussion in Section \ref{disk_structure}).

In order to derive the magnetic torque $T_{\rm m}$, we need to calculate the mass loss rate in the outflow. It is well known that the mass loss rate is regulated by the density at the sonic point. The location of the sonic point can be evaluated through the Bernoulli equation of the outflow along a certain field line,
\begin{equation}
{\frac 1 2}v_{\rm w}^2+c_{\rm s,s}^2\ln \rho_{\rm w}+\Psi_{\rm eff}=const,\label{bernoulli}
\end{equation}
which shows that it is always around the point corresponding to the maximal effective potential \citep*[see, e.g.,][for the details]{2019ApJ...872..149L}. As the effective potential is given by
\begin{equation}
\Psi _{\rm eff}(R,z) = - {\frac {GM_{\rm BH}}{( R^2 + z^2)^{1/2}}} - {\frac{1}{2}}R^2 \Omega ^2, \label{psi}
\end{equation}
the relation of the gas density at the sonic point is related to that at the disk surface by
\begin{equation}
\rho_{\rm w,s}\simeq \rho_{\rm w,i}\exp\left({\frac {\Psi_{\rm eff,s}-\Psi_{\rm eff,i}}{c_{\rm s,s}^2}}\right)\simeq \rho_{\rm w,i}\exp\left({\frac {\Psi_{\rm eff,max}-\Psi_{\rm eff,i}}{c_{\rm s,s}^2}}\right),\label{rho_w_s}
\end{equation}
where $\rho_{\rm w,i}$ is the gas density at the disk surface, and Equation (\ref{bernoulli}) is used. In principle, this can be calculated self-consistently by including the physics taking place in the transition region between the disk and outflow, which has already been studied by the previous works, however, some additional parameters have to be induced to describe the problem \citep*[][]{1998ApJ...499..329O,2001ApJ...553..158O}. This is beyond the scope of this work. In this work, we use a parameter $\beta_{\rm s}$ to estimate the gas density at the base of the outflow,
\begin{equation}
\beta_{\rm s}={\frac {8\pi P_{\rm gas,i}}{B_{\rm p,i}^2}}={\frac {8\pi\rho_{\rm w,i}c_{\rm s,s}^2}{B_{\rm p,i}^2}}.\label{beta_s}
\end{equation}
It is certain that the gas pressure at the disk surface (the base of the outflow) must be lower than the magnetic pressure (i.e., $\beta_{\rm s}<1$), so that the gas at the disk surface can be driven into the outflow efficiently.
Thus, the mass loss rate in the outflow from unit area of the disk surface is estimated as
\begin{equation}\label{mass_load_para_solve-2}
\dot{m}_{\rm w}=\rho_{\rm w}c_{\rm s,s}\simeq
\frac{\beta_{\rm s }B_{\rm p,i}^2}{8 \pi c_{\rm{s,s}}}
 {\rm exp} \left( -\frac{\Psi _{\rm{eff,s}} -
 	\Psi _{\rm{eff,0}} }{c^2_{\rm{s,s}}}  \right),
\end{equation}
where Equations (\ref{rho_w_s})-(\ref{mass_load_para_solve-2}) are used. The magnetic torque $T_{\rm m}$ exerted on the disk can be calculated with Equations (\ref{T_m_nu}), (\ref{mass_load_para}), and (\ref{mass_load_para_solve-2}), when the field strength, the temperature and density of the gas at the surface of the disk are known.

\begin{figure}
\centering
\includegraphics[width=16cm,clip=]{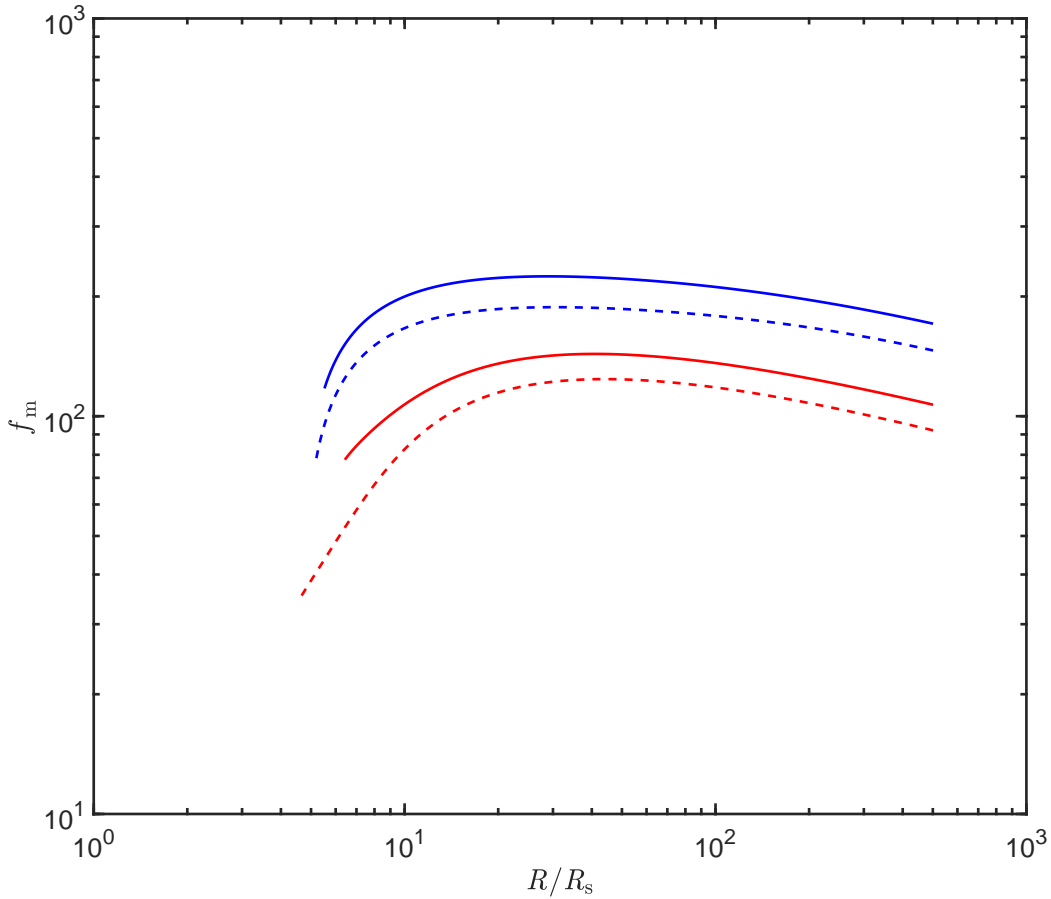}
\caption{The amplification of the radial velocity of the disks with outflows varies with radius. The red lines represent the disk with an accretion rate at outer boundary,  $\dot{m}_{\rm out}=0.5$, while the blue lines represent the disk accreting at $\dot{m}_{\rm out}=0.05$. The results calculated with $\alpha = 0.1$ are showed as dashed lines, while the solid lines represent the results with $\alpha = 0.5$.}
\label{fig:fm}
\end{figure}


\begin{figure}
\centering
\includegraphics[width=16cm,clip=]{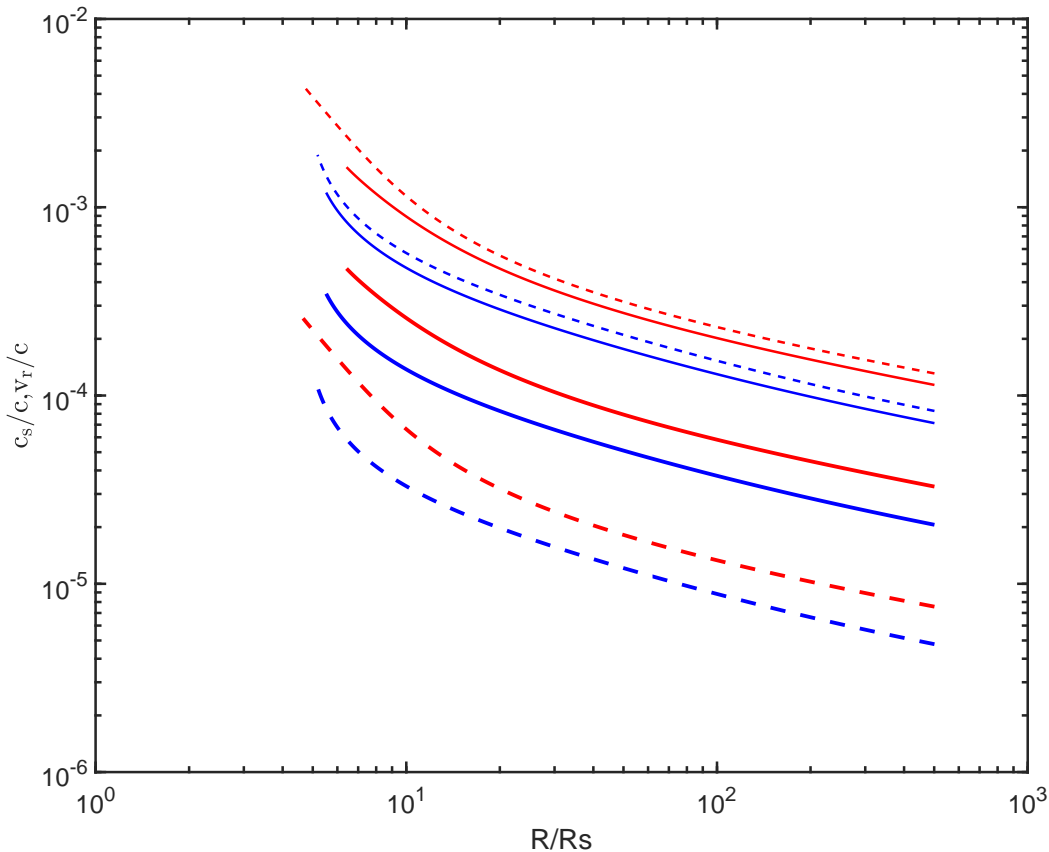}
\caption{The radial velocity and sound speed in speed of light vary with radius calculated with different values of  $\dot{m}_{\rm out}$ and $\alpha$ (the thick lines). The thin lines represent the sound speeds. The values of all input model parameters are the same as Figure \ref{fig:fm}.}
\label{fig:vrcs}
\end{figure}


\begin{figure}
\centering
\includegraphics[width=16cm,clip=]{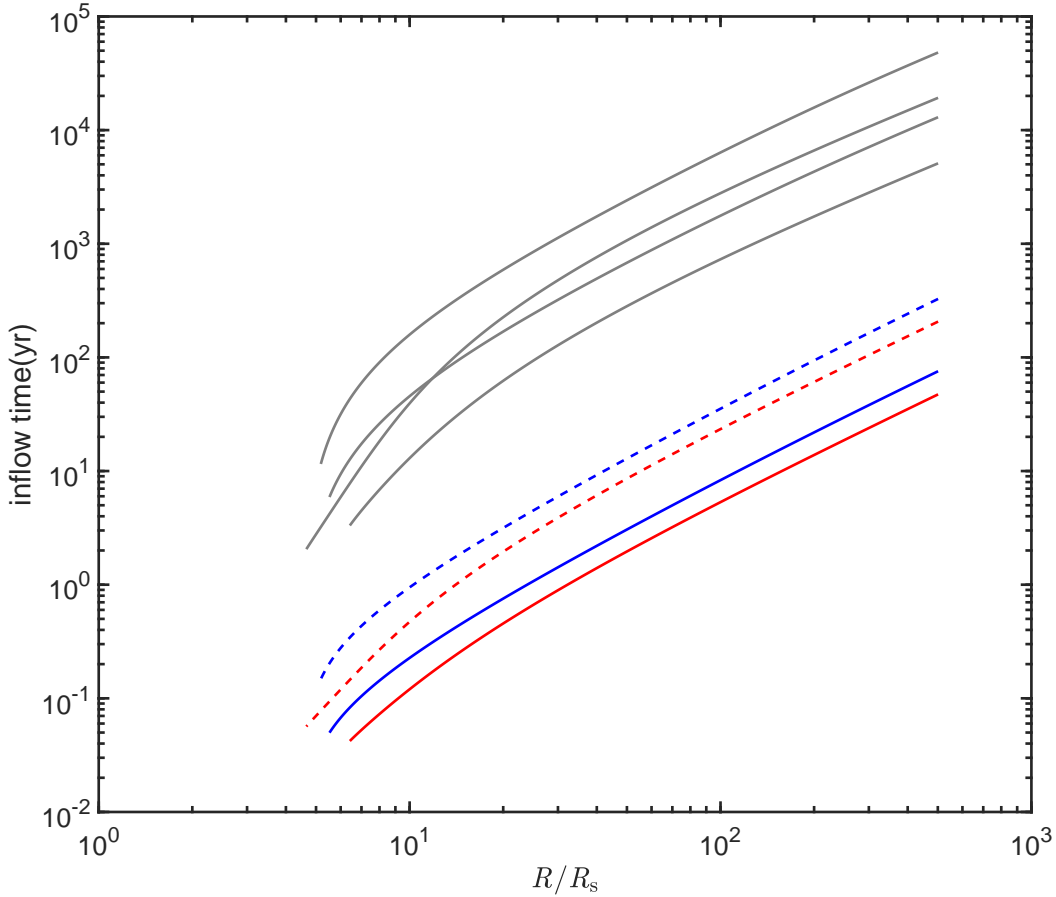}
\caption{The inflow time varies with radius calculated with different values of  $\dot{m}_{\rm out}$ and $\alpha$. The gray solid lines are viscous time-scale calculated by standard thin disk model. The values of all input model parameters are the same as Figure \ref{fig:fm}. }
\label{fig:time}
\end{figure}


\begin{figure}
\centering
\includegraphics[width=16cm,clip=]{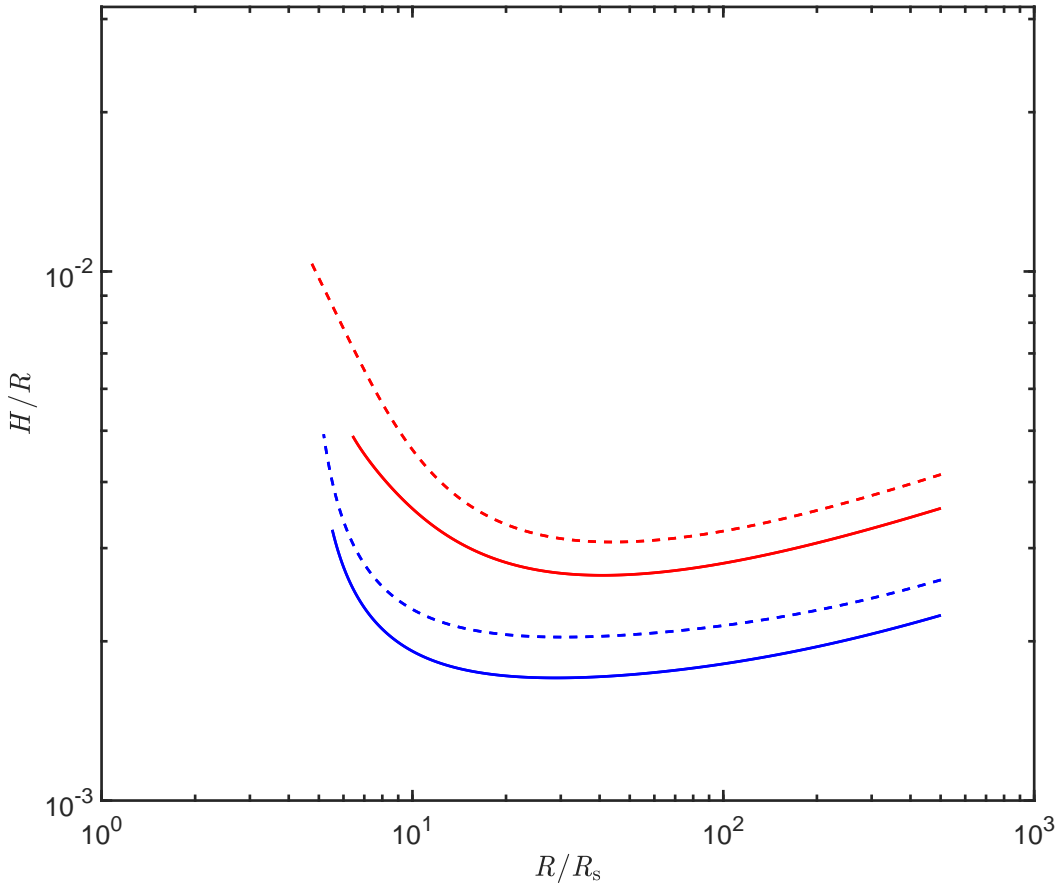}
\caption{The scale height varies with radius. The values of all input model parameters are the same as Figure \ref{fig:fm}.}
\label{fig:h}
\end{figure}


\begin{figure}
\centering
\includegraphics[width=16cm,clip=]{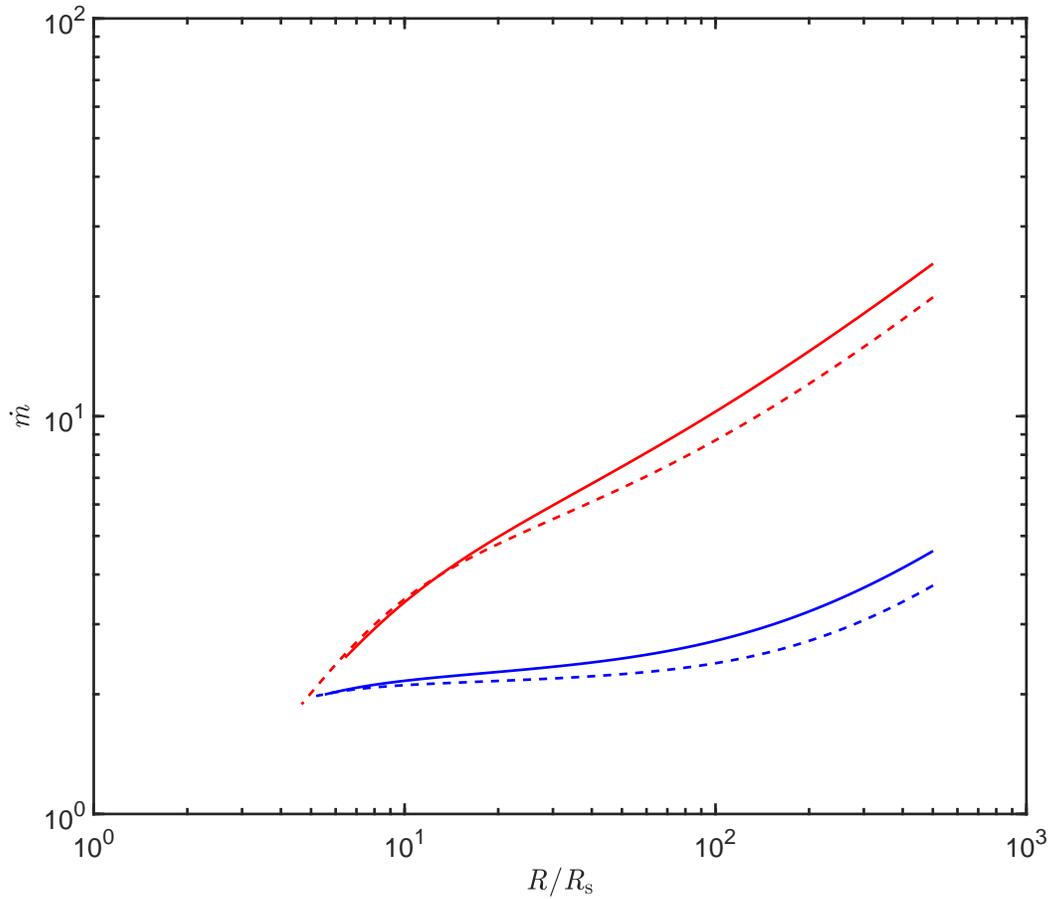}
\caption{The accretion rate varies with radius. The values of all input model parameters are the same as Figure \ref{fig:fm}. }
\label{fig:mdot}
\end{figure}


\begin{figure}
\centering
\includegraphics[width=16cm,clip=]{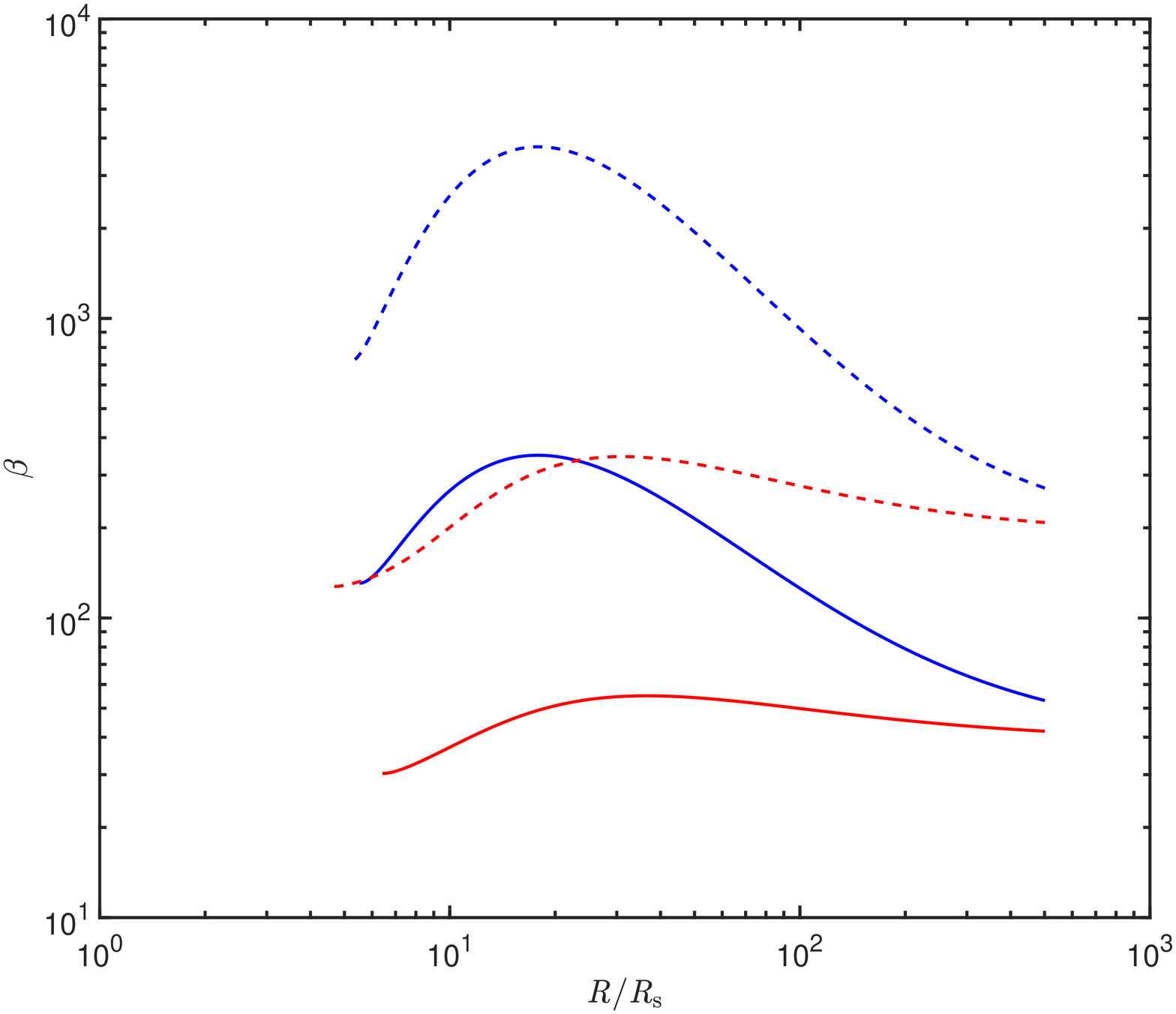}
\caption{The strength of the large-scale magnetic field varies with radius. The values of all input model parameters are the same as Figure \ref{fig:fm}.}
\label{fig:beta}
\end{figure}

\section{Numerical method} \label{sec:results}

In this model, most angular momentum of the gas in the disk is carried away by the magnetically driven outflows. It is well known that such outflows require the field lines to be inclined to the disk surface at an angle $\la 60^\circ$ for a thin standard accretion disk \citep*[][]{1982MNRAS.199..883B}. The calculations of the field configuration of such an thin disk with outflows are rather complicated \citep*[][]{2013ApJ...765..149C,2019ApJ...872..149L,2021ApJ...909..158L}. In this work, we fix $\kappa_0=\sqrt{3}$ in all of the calculations for simplicity. There are three input model parameters, central BH mass $M_{\rm BH}$, viscosity parameter $\alpha $, gas density at the bottom of the outflow $\beta_{\rm s}$, and the magnetic Prandtl number $ {\cal{P}}_{\rm m} $.
The calculation is started from the outer radius $R_{\rm out}$ of the disk with mass accretion rate $\dot{m}_{\rm out}$ ($\dot{m}_{\rm out}=\dot{M}_{\rm out}/\dot{M}_{\rm Edd}$).
 The disk is closely coupled with outflows through angular momentum and mass transfer between the disk and outflows. With specified mass accretion rate $\dot{M}(R)$, we have to solve 9 algebraic equations (Equations \ref{mdot}, \ref{v_R}, \ref{diff_omega}, \ref{H_over_R}, \ref{radiation}, \ref{v_sound}, \ref{f_m}, \ref{T_s}, and \ref{T_m_nu}) for 9 variables, $\rho$, $T_{\rm c}$, $T_{\rm s}$, $B_z$, $c_{\rm s}$, $H$, $\Omega$, $T_{\rm m}$, and $\dot{m}_{\rm w}$.

  Integrating the continuity equation (\ref{dM_acc-dR}) from the outer radius $R_{\rm out}$ inwards, the mass accretion rate as a function of $R$ is available with the derived mass loss rate $\dot{m}_{\rm w}(R)$ using the method described above. In this way, the whole structure of the disk with outflows is available.

\section{Results} \label{sec:results_b}

In all of the calculations, we adopt a typical value of ${\cal{P}}_{\rm m}=1$ \citep*[][]{1979cmft.book.....P,2003A&A...411..321Y,2009A&A...504..309L,2009A&A...507...19F,
2009ApJ...697.1901G}. The local structure of the accretion disk-outflow system is calculated from $R=500R_{\rm s}$ ($R_{\rm s}=2GM_{\rm BH}/c^2$), when the values of $\alpha$ and $\dot{M}(R)$ are specified.



First, we calculate the structure of the disk with outflows for a typical super-massive BH with $M_{\rm BH}=10^7M_{\odot}$. The calculations are carried out with four sets of the parameter values, i.e.,  a). $\beta_{\rm s}=0.1$, $\alpha=0.5$, and $\dot{m}_{\rm out}=0.5$; b). $\beta_{\rm s}=0.1$, $\alpha=0.5$, and $\dot{m}_{\rm out}=0.05$; c). $\beta_{\rm s}=0.5$, $\alpha=0.1$, and $\dot{m}_{\rm out}=0.5$; d). $\beta_{\rm s}=0.5$, $\alpha=0.1$, and $\dot{m}_{\rm out}=0.05$, respectively.

Figure \ref{fig:fm} shows how the amplification $f_{\rm m}$ of the radial velocity of the disks with outflows varies with the radius for different values of $\alpha$ and $\dot{m}_{\rm out}$. We plot the radial velocities and the sound speeds with different values of the model parameters in Figure \ref{fig:vrcs}. 
We find that the radial velocities of the disks with magnetic outflows can increase to hundreds to thousands times higher than those of the conventional viscous disks, however,  they are still sub-sonic (see the thin lines in Figure \ref{fig:vrcs}). This is different from the magnetic accretion disk-outflow model developed for the X-ray binaries \citep{2006A&A...447..813F,2018A&A...615A..57M,2018A&A...617A..46M,2019A&A...626A.115M,2020A&A...640A..18M}. 

The inflow time is then calculated with $t_{\rm inflow}\sim -R/v_R$ and Equation (\ref{v_R}), and we plot the results in Figure \ref{fig:time}. We also plot the viscous time-scale of a standard thin disk with the same values of the parameters. We find that the inflow time of the disk with outflows is much shorter than that of a standard thin disk. In all of our calculations, the inflow time is less than 20 years at $R=100 R_{\rm s}$, where is the typical region of the disk emitting UV/optical photons in AGNs.

\begin{figure}
\centering
\includegraphics[width=16cm,clip=]{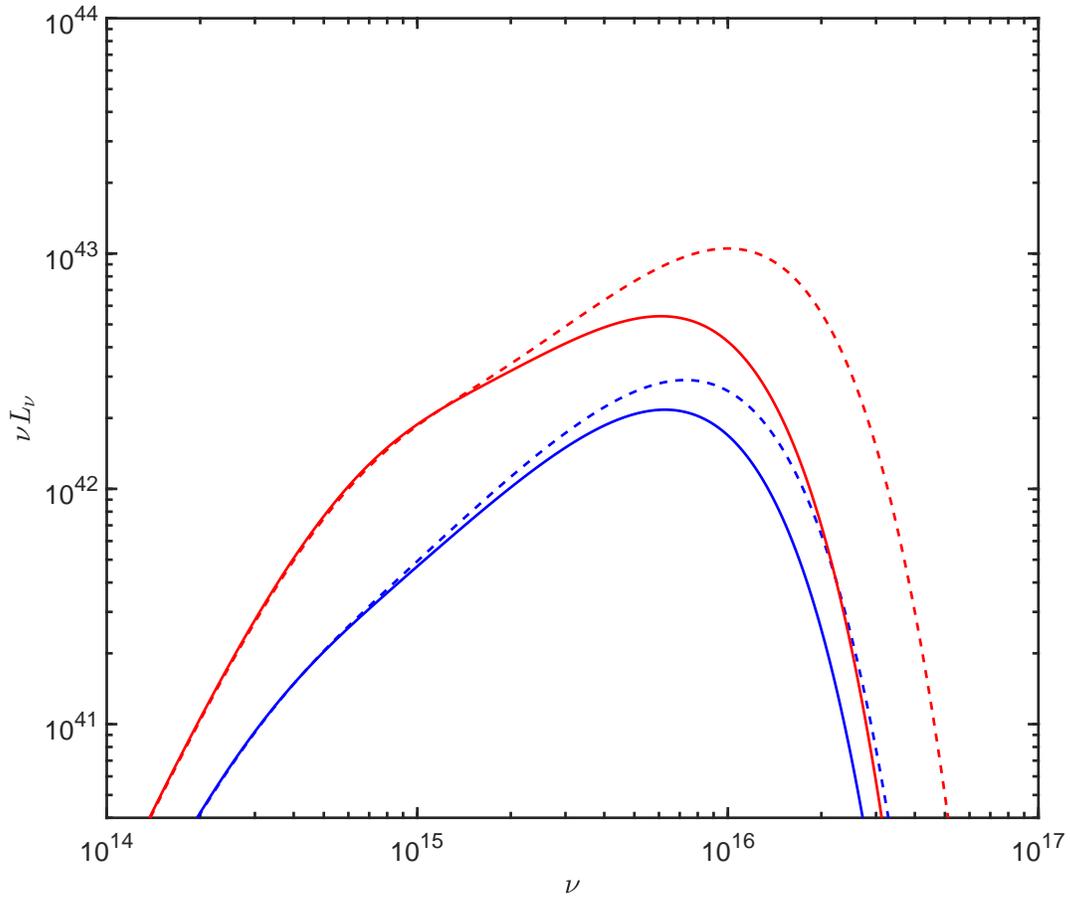}
\caption{The continuum spectra of the disk with magnetic outflows. The values of all input model parameters are the same as Figure \ref{fig:fm}. }
\label{fig:spectra}
\end{figure}

Figure \ref{fig:h} shows the scale height varies with the radius. We find that the disk with outflows always remains geometrically thin even if the mass accretion rate at the outer radius of the disk $\dot{m}_{\rm out}\gg0.2$, which is due to the fact that a large fraction of the gravitational power of the gas released is taped into the outflows by the magnetic field co-rotating with the disk. The radiation of the disk is substantially sub-Eddington even when $\dot{m}_{\rm out}\gg 1$ (see Figure \ref{fig:spectra}). Our calculations show that Eddington ratio are $L/L_{\rm Edd}=0.0269$, $0.0098$, $0.0488$ and $0.0129$, corresponding to the cases of a, b, c, and d.

For the disk with magnetically driven outflows, its accretion rate decreases with decreasing radius (see Figure \ref{fig:mdot}). We find that a large fraction of mass supplied at the outer radius of the disk is fed into the outflows, and only a small fraction of the mass is swallowed by the BH.

\begin{figure}
\centering
\includegraphics[width=16cm,clip=]{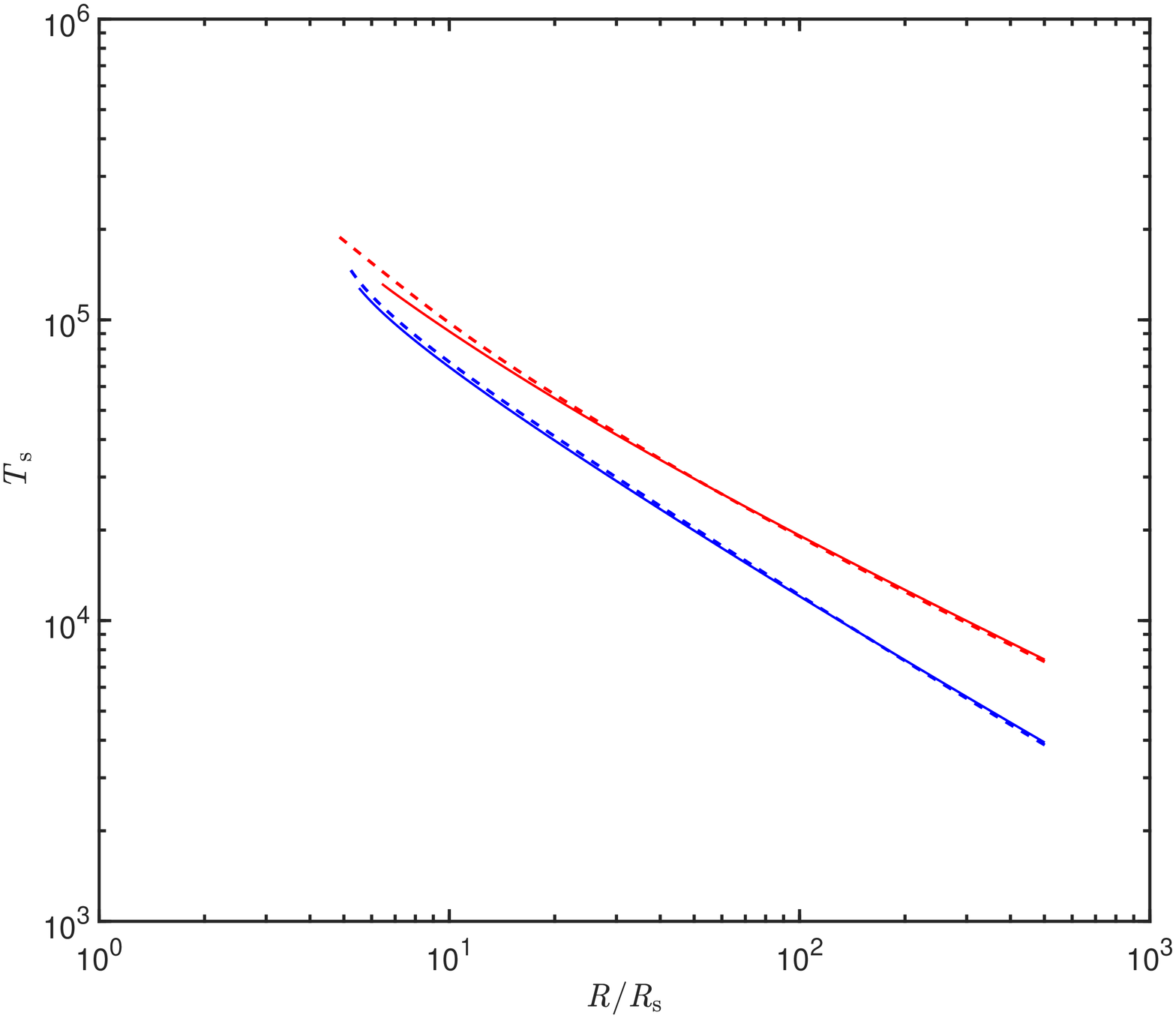}
\caption{The surface temperature varies with radius. The values of all input model parameters are the same as Figure \ref{fig:fm}.}
\label{fig:Te}
\end{figure}

\begin{figure}
\centering
\includegraphics[width=16cm,clip=]{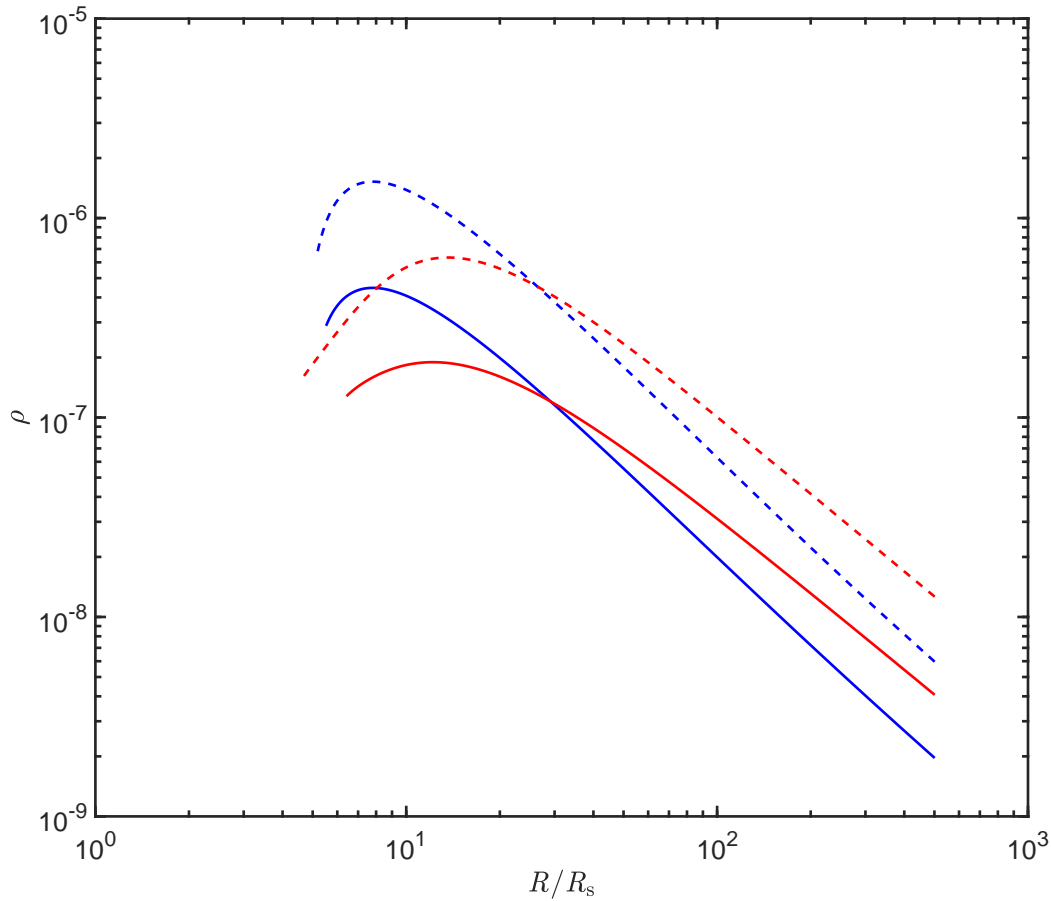}
\caption{The mass density varies with radius. The values of all input model parameters are the same as Figure \ref{fig:fm}.}
\label{fig:rho}
\end{figure}

In Figure \ref{fig:beta}, we plot the ratio of the gas pressure to the magnetic pressure of the disk. It is found that a moderate field with $\beta\sim 10-100$ is required in such an accretion disk with outflows ($\beta \sim 100$ in the outer region of the disk with $\alpha=0.1$, and $\beta \sim 10$ for $\alpha=0.5$), which justifies the assumption of the vertical compression of the disk by the magnetic field being negligible. We also plot the disk structure in Figure \ref{fig:Te}-\ref{fig:rho}.

Our calculations indeed show that the inflow time can be as short as several years. Although the physical mechanism responsible for CL AGNs is still unclear
\citep[e.g.,][]{2017MNRAS.467.1496O,2020ApJ...890L..29A,2015MNRAS.452...69M}, we can tentatively explain the variability of CL AGNs as it being triggered by mass accretion rate changes in the disk with magnetic outflows.

\begin{figure}
\centering
\includegraphics[width=16cm,clip=]{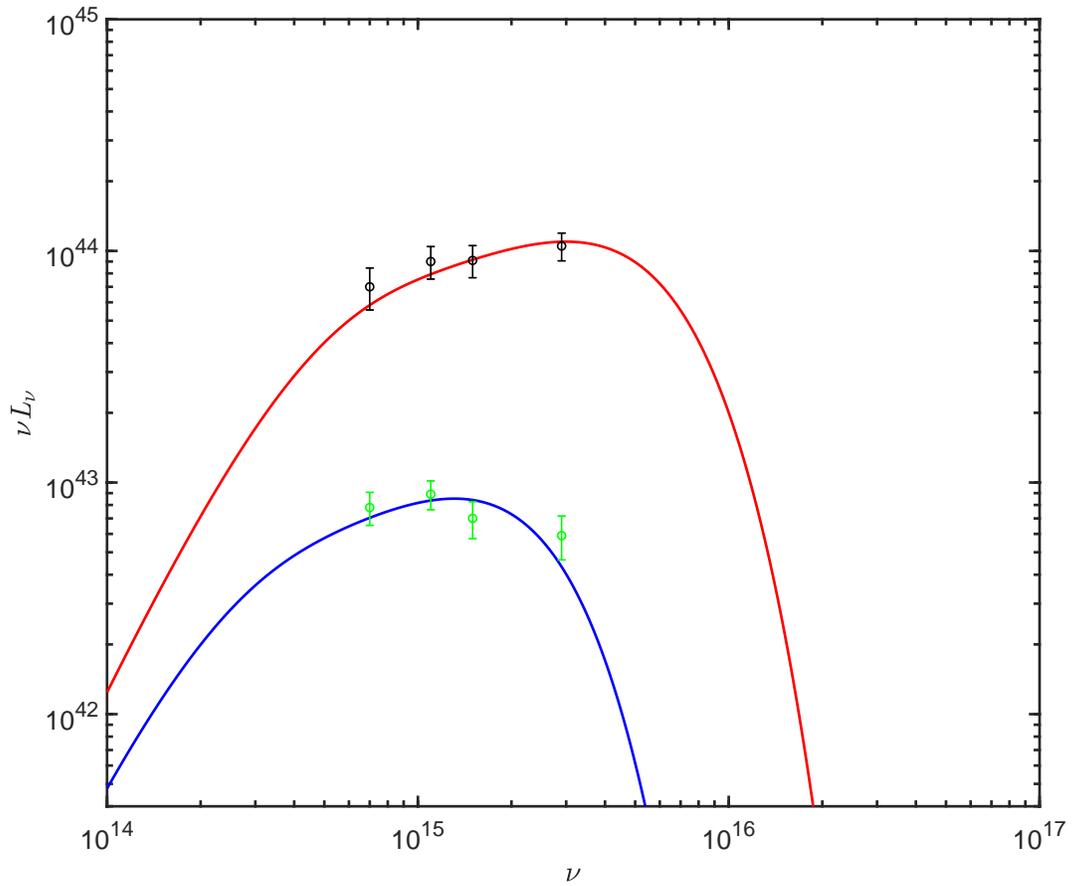}
\caption{The continuum spectra of the disk with magnetic outflows accreting at different rates. The black dots represent the observation of Mrk 1018 in Feb. 2008 (high state), and the green dots represent the observation of Mrk 1018 in Feb 2016 (low state) \citep[see][for details]{2016A&A...593L...9H}. The red line represents the disk with an accretion rate at outer boundary,  $\dot{m}_{\rm out}=1.78$, while the blue line represents the disk accreting at $\dot{m}_{\rm out}=0.13$. The black hole mass $M_{\rm BH}=10^8 M_{\odot}$ is adopted in this calculation. }
\label{fig:m8spectra}
\end{figure}
\begin{figure}
\centering
\includegraphics[width=16cm,clip=]{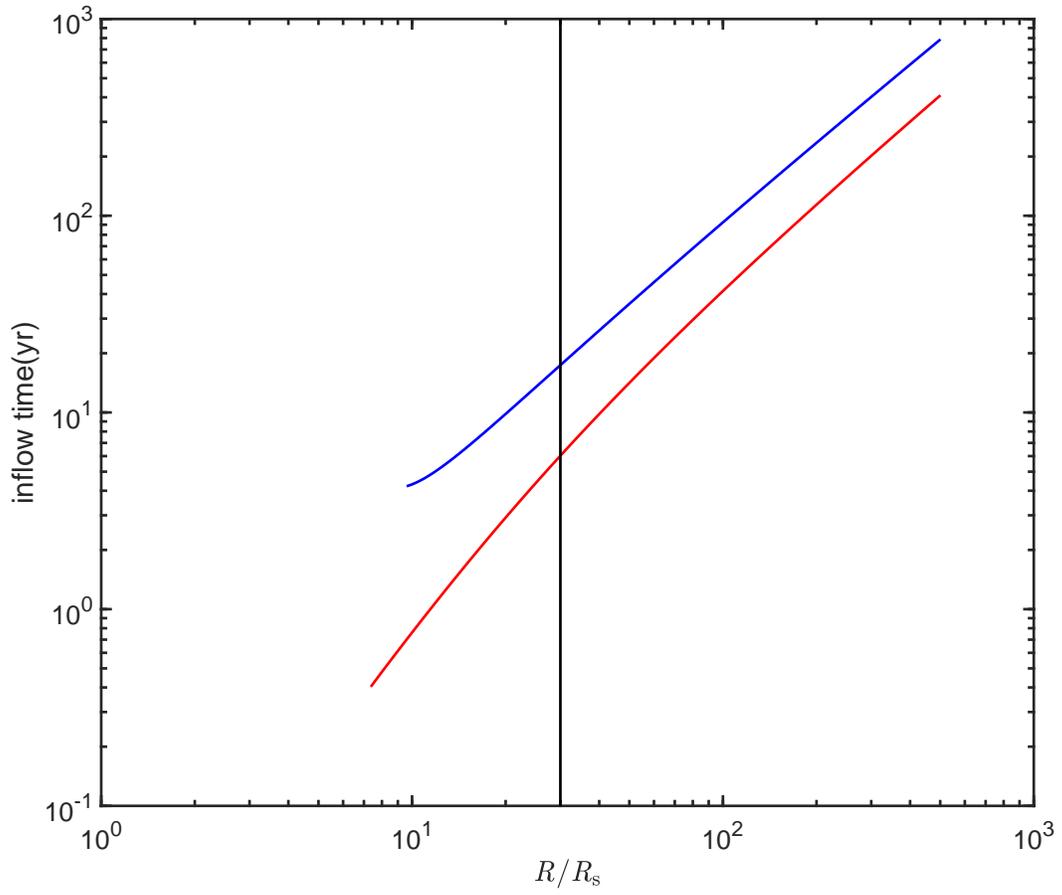}
\caption{The inflow time varies with radius applied to Mrk 1018. The red line represents the disk with an accretion rate at outer boundary,  $\dot{m}_{\rm out}=1.78$, while the blue line represents the disk accreting at $\dot{m}_{\rm out}=0.13$. The black hole mass $M_{\rm BH}=10^8 M_{\odot}$ is adopted in this calculation.}
\label{fig:m8time}
\end{figure}

We further apply our model to a well observed CL AGN Mrk 1018($z=0.035$, of which the BH mass is estimated as $log(M_{\rm BH}/M_{\odot})=7.4-7.9$. The Eddington ratios are $L/L_{\rm Edd}=0.03$ and $L/L_{\rm Edd}=0.004$ in the high and low states, respectively \citep[see][and the references therein]{2016A&A...593L...8M,2016A&A...593L...9H,2018MNRAS.480.3898N}. Mrk 1018 changed from type 1.9 to type 1 in 5 years \citep{1986ApJ...311..135C,1989ApJ...340..190G}. \citet{2016A&A...593L...8M}'s observations show that Mrk 1018 changed from type 1 to type 1.9 after about 30 years.
We adopt the black hole mass $M_{\rm BH}=10^8 M_{\odot}$ in our model calculations for Mrk 1018. The spectra of the disk with outflows are able to reproduce the observed optical/UV spectra in the low/high states quite well (observed in Feb. 2008 and Feb. 2016, respectively) in Figure \ref{fig:m8spectra}, if $\alpha=0.5$, $\beta_{\rm s}=0.4$ and the mass accretion rates, $\dot{m}_{\rm out}=0.13$ and $\dot{m}_{\rm out}=1.78$ corresponding to low and high state respectively. The inflow time of the disks with best fits to the observations at low/high states is plotted in Figure \ref{fig:m8time}, as the UV/optical photons are emitted from the disk region around $R=30R_{\rm s}$. Our calculations show that the inflow time is about 20 years for low mass accretion rate cases, and could be less than 5 years when the mass accretion rate $\dot{m}$ is high, which is consistent with the observed timescale of CL AGN Mrk 1018.

\begin{figure}
\centering
\includegraphics[width=16cm,clip=]{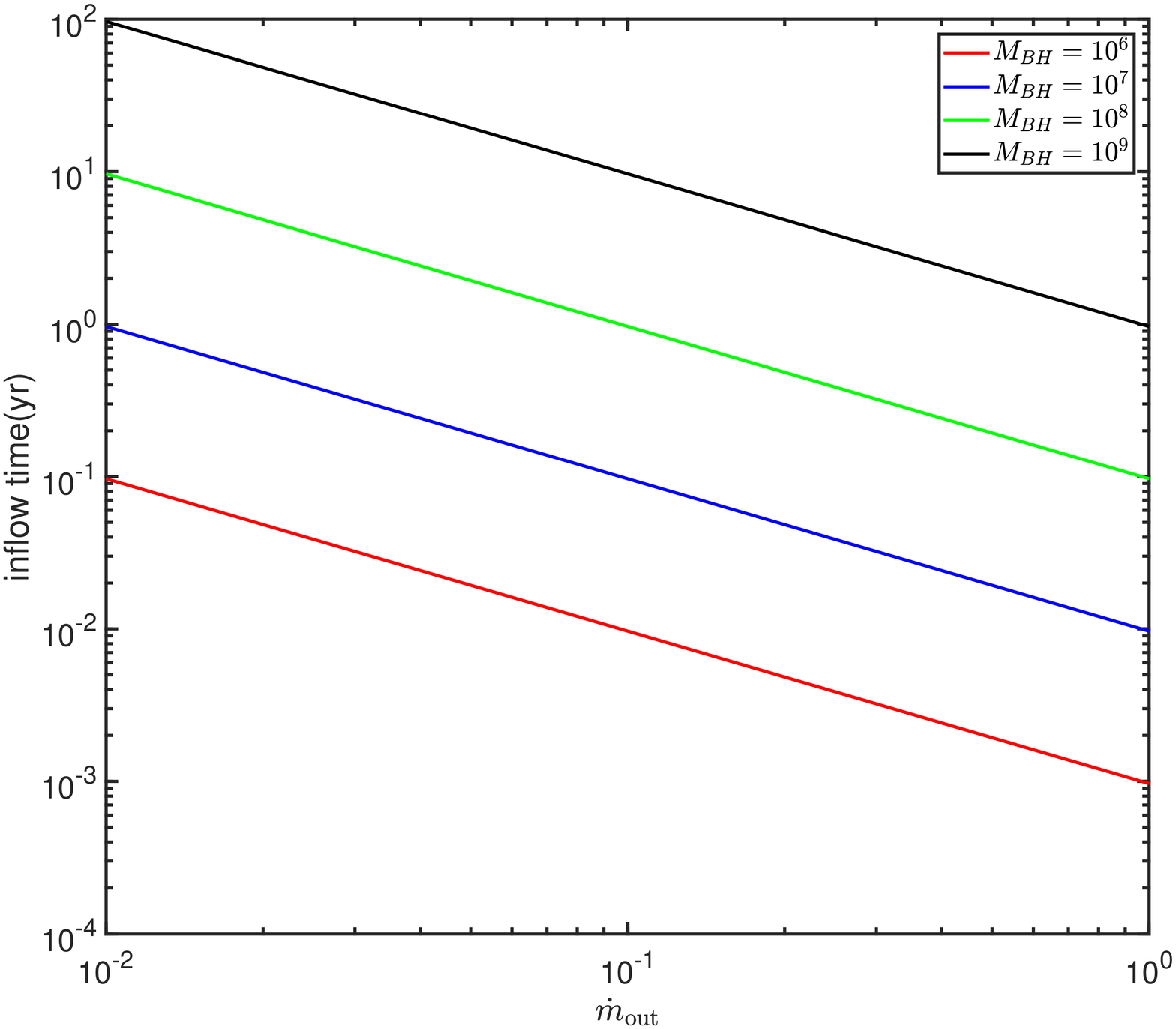}
\caption{The inflow times at $R=30R_s$ vary with $\dot{m}_{\rm out}$ ($\alpha=0.5$ is adopted).}
\label{fig:limit}
\end{figure}

In all previous calculations, a fixed black hole mass $M_{\rm BH}=10^7M_{\odot}$ is adopted. We plot the inflow time of the disk with magnetic outflows varying with the black hole mass in Figure \ref{fig:limit} ($\alpha=0.5$ is adopted).  


\section{Discussion and Conclusions} \label{sec:discussion}

For a standard thin accretion disk, the radiation in UV/optical bands may be from the inner region of the disk (around several tens to one hundred Schwarzschild radii for a $10^8M_\odot$ BH accreting at a moderate rate). The viscous timescale at this radius could be as high as $\sim 10^5$ years. It is obvious that the changes of mass accretion rate in a standard thin disk is unable to reproduce the observed variabilities in CL AGNs, of which the timescales are much shorter \citep*[see][and the references therein]{2018MNRAS.480.3898N,2019MNRAS.483L..17D,2020A&A...641A.167S}.

In this work, we use a magnetic disk-outflow model developed by \citet{2013ApJ...765..149C} to calculate the local structure of the disk with magnetically  driven outflows. With derived local disk and outflow structure, the whole disk structure is available with integrating the continuity equation from the outer radius of the disk. A large fraction of the angular momentum and gravitational energy released in the disk is carried away by the magnetic outflows, which substantially increases the radial velocity of the disk. This naturally shortens the inflow time of the disk, also reduces the disk radiation significantly, i.e., it could be sub-Eddington even if the dimensionless accretion rate at the outer region of the disk is substantially higher than the unity.

Using typical values of the model parameters, we find that the moderate field strength may lead to significantly increased radial velocity of the disk with outflows (see Figure \ref{fig:beta}), which decreases the inflow time substantially to the order of years (see Figure \ref{fig:time}). We tentatively apply the magnetic accretion disk-outflow model to a CL AGN Mrk 1018, and we find that the observed UV/optical spectra both in low and high state can be fairly well fitted by our model calculations (see Figure \ref{fig:m8spectra}). In our model, the continuum spectra of the disks with outflows are calculated based on the  multi-color black body assumption, which have not included the reradiation from the putative dust torus. This may lead to deviation in the infrared band. The detailed spectral modelling of this source including the dust emission is beyond the scope of the present work.


Our calculations show that the inflow time of the disk is $\sim 5$ years in the high accretion state, while it becomes $\sim 20$ years in the low accretion state. This is consistent with the observational features of Mrk 1018. It indicates that the different CL time-scales between type 1 to type 1.9 transition and type 1.9 to 1 transition can be caused by variation of the accretion rate. We note that the physical processes triggering mass accretion rate variation have not been included in our model calculations, which might be the focus of the future work.

The repeated CL events may have very short periods (e.g., X-ray quasi-periodic eruptions in AGNs) as short as 9 hours \citep{2019Natur.573..381M,2020A&A...636L...2G,2021Natur.592..704A}. We calculated how the inflow time varies with at $R=30R_s$ in our calculations with the BH mass and accretion rate (see Figure \ref{fig:limit}). We find that the inflow time is always higher than 9 hours even with a sufficiently high mass accretion rate and a low BH mass, which implies that the simple mass accretion variation in an accretion disk with magnetic outflows is unable to reproduce the features in QPEs (at least for those with short periods). We conjecture that the instability model of an accretion disk with magnetic outflows may help explain the QPEs in AGNs \citep*[][and Pan, Li, \& Cao in preparation]{2021ApJ...910...97P}, which is beyond the scope of this work. 


\acknowledgments

We thank the referee for helpful comments/suggestions. This work is supported by the NSFC (11773050, 11833007, 12073023, and 12033006), the CAS grant QYZDJ-SSWSYS023.

%





\bibliography{CLAGN}{}

\bibliographystyle{aasjournal}



\end{document}